\documentclass[
reprint,
superscriptaddress,
bibnotes,
amsmath,
amssymb,
aps,
floatfix,
]{revtex4-2}

\usepackage{xcolor,mathrsfs,dsfont, float}
\definecolor{darkblue}{RGB}{0,0,150}
\definecolor{nightblue}{RGB}{0,0,100}

\usepackage{graphicx,mathtools,bm}
\usepackage[
colorlinks,
citecolor=darkblue,
linkcolor=darkblue,
urlcolor=nightblue]{hyperref}

\pdfoutput=1

\usepackage[english]{babel}
\usepackage[babel,kerning=true,spacing=true]{microtype}
\usepackage{tabulary}
\usepackage{tabularx, siunitx}
\usepackage{appendix,stmaryrd}

\newcommand{\refsub}[2]{\hyperref[#1]{\ref*{#1}#2}}

\bibpunct{[}{]}{,}{n}{}{}
\makeatletter
\def\NAT@def@citea{\def\@citea{\NAT@separator}}
\makeatother

\begin{document}

	\title{
		Connecting cooperative transport by ants with the physics of self-propelled particles
	}
	\author{Tabea Heckenthaler}
	\email{tabea.heckenthaler@weizmann.ac.il}
	\affiliation{Department of Physics of Complex Systems, Weizmann Institute of Science, Rehovot, Israel 76100}
	\author{Tobias Holder}
	\thanks{T.H. and T.H. have contributed equally to this work.}
	\affiliation{Department of Condensed Matter Physics, Weizmann Institute of Science, Rehovot, Israel 76100}
	\author{Ariel Amir}
	\affiliation{Department of Physics of Complex Systems, Weizmann Institute of Science, Rehovot, Israel 76100}
	\affiliation{John A. Paulson, School of Engineering and Applied Sciences, Harvard University, Cambridge, Massachusetts 02138, USA}
	\author{Ofer Feinerman}
	\affiliation{Department of Physics of Complex Systems, Weizmann Institute of Science, Rehovot, Israel 76100}
	\author{Ehud Fonio}
	\affiliation{Department of Physics of Complex Systems, Weizmann Institute of Science, Rehovot, Israel 76100}
	\date{\today}
	
	\begin{abstract}
		\textit{Paratrechina longicornis} ants are known for their ability to cooperatively transport large food items.
		Previous studies have focused on the behavioral rules of individual ants and explained the efficient coordination using the coupled-carriers model.
		In contrast to this microscopic description, we instead treat the transported object as a single self-propelled particle characterized by its velocity magnitude and angle.
		We experimentally observe \textit{P. longicornis} ants cooperatively transporting loads of varying radii. By analyzing the statistical features of the load's movement, we show how the salient properties are encoded in its deterministic and random accelerations which are the basis of our model. We relate the parameters of our macroscopic model to the parameters of the previous coupled-carriers model. While the autocorrelation time of the velocity direction increases with group size, the autocorrelation time of the speed has a maximum at an intermediate group size. This corresponds to the critical slowdown close to the phase transition identified in the coupled-carriers model. 
		Our findings illustrate that a self-propelled particle model can effectively characterize a system of interacting individuals.
	\end{abstract}
	
	\maketitle
	
	\section{Introduction}
	Cooperative transport by ants is the concerted effort to carry large food items toward their nest. This behavior is observed in a variety of ant species~\cite{mccreery2014, czaczkes2011, mccreery2016}. Longhorn crazy ants (\textit{Paratrechina longicornis}) display some of the most impressive cooperative transport abilities. These two-and-a-half millimeter ants gather in groups consisting of hundreds of individuals to haul loads that are heavy and large, exceeding 10,000 times their own weight and more than one hundred times their body length~\cite{Gelblum2016}. Moreover, these ants show remarkable collective navigation abilities as they transport large loads across complex obstacles~\cite{Gelblum2016, Ayalon2021, mccreery2016} and through disordered environments~\cite{Gelblum2020} to deliver them to their nest quickly. 
	To achieve this efficiency and robustness, the ants rely on the following cooperative behavior:
	When a single ant finds a food item which is too large for her to transport by herself to her nest, she lays a pheromone trail in order to recruit ants from the nest to the site~\cite{Fonio2016}. Once enough ants are gathered, they cooperate to transport the item using their mandibles, assuming various roles during transport: Newly attached ants function as temporary leaders as they are informed of the nest location. They persist in pulling the load towards the nest irrespective of the current direction of motion for approximately ten seconds~\cite{Feinerman2017,Boczkowski2018,Ayalon2021}. Ants that have been connected to the load for more than ten seconds sense the current direction of motion and align their efforts accordingly. They pull the load if they are attached at the leading edge and lift the load to reduce friction with the floor if they are attached to the trailing edge ~\cite{Gelblum2015}.
	There is a constant turnover of ants, as carriers sometimes let go of the load, and unattached ants take their place. \\
	\begin{figure}[b]
		\centering
		\includegraphics[width=\columnwidth]{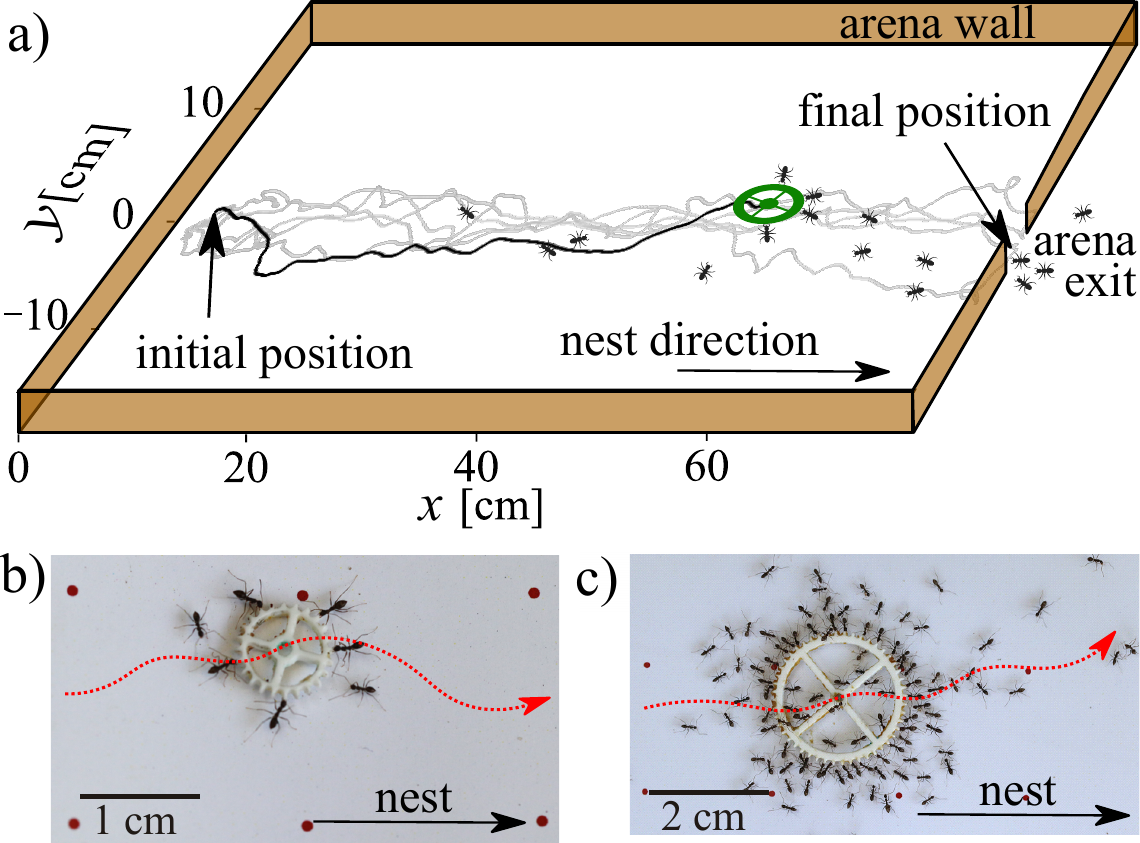}
		\caption{Cooperative transport trajectories. a) Experimental trajectories (gray) of transported silicon ring with radius \textcolor{black}{$r=\SI{0.4}{cm}$}. One trajectory is highlighted in black with an illustration of ants carrying the object. Photographs of groups of ants cooperatively transporting rings of radii $r = \SI{0.4}{cm}$ and $\SI{1.0}{cm}$, respectively. The red dotted lines illustrate the trails are along which the centers of the rings move towards the nest.}
		\label{fig:1}
	\end{figure}
	These relatively simple behavioral rules are the basis of the agent-based coupled carriers model~\cite{Feinerman2017, Boczkowski2018, Ayalon2021, Feinerman2018, Gelblum2015, Gelblum2016}.
	Numerical simulations within this model's framework revealed a critical finite-size phase transition between uncoordinated (individualistic) and coordinated ant behavior that sets an ideal group size which maximizes the collective response to an informed ant~\cite{Gelblum2015,Gelblum2016,Feinerman2018,Ron2018}. While this model successfully replicates load trajectories for different group sizes, it does not provide a coarse-grained, analytic understanding of load trajectories.
	
	In this paper, we show how a self-propelled particle model~\cite{Romanczuk2012,Marchetti2013,Elgeti2015,Shaebani2020} can be used to quantitatively describe the macroscopic features of the carried load's motion. To this end, we experimentally observed trajectories of spatially unconstrained rings of varying radii which are carried by \textit{P. longicornis} ants (\textit{cf.} Fig.~\ref{fig:1}) and developed a statistical description of these trajectories. 
	We show that a model originally developed to describe cell chemotaxis~\cite{Schienbein1993} can be employed to describe ants that are transiently tethered together by a load as one large self-propelled particle. To our knowledge, cooperative transport presents the first example of collective motion in an animal group which can be characterized by the physics of a single effective agent~\cite{Vicsek2012,Hu2016}. 
	
	In the following, we first present the experiment and discuss the statistical properties of the trajectory data. Then, we introduce a four-parameter description of the data in terms of deterministic and random accelerations of the velocity's angle and magnitude. Finally, we analyze the resulting model parameters as a function of the ring size which correlates with the number of carrying ants linking the different scaling behaviors to the phase transition demonstrated in the coupled-carriers model.
	
	\section{Cooperative transport experiment}
	
	\begin{figure*}[h]
		\centering
		\includegraphics[width=\linewidth]{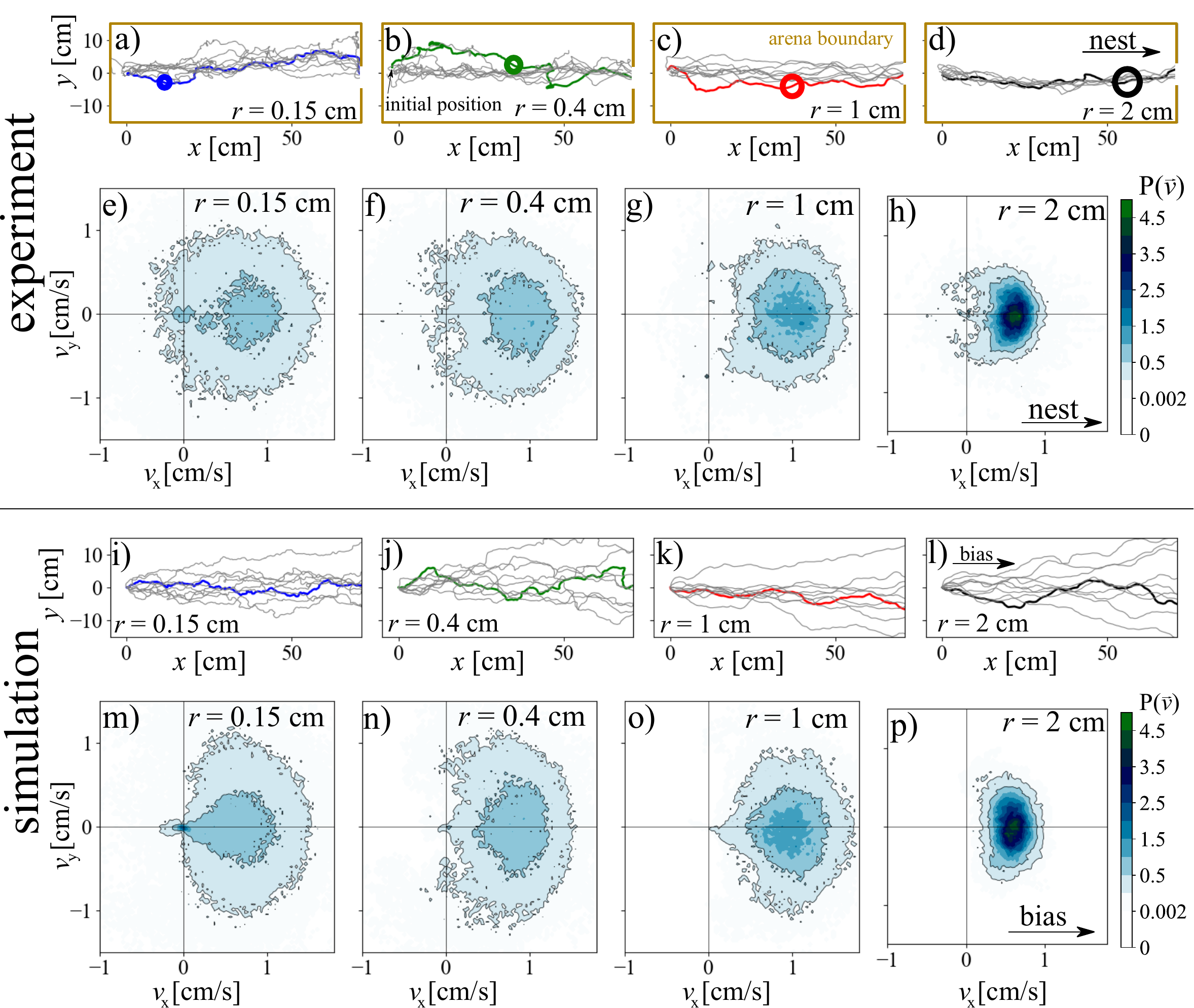}
		\caption{Trails of the center of rings transported by ants with radii $r=\SI{0.15}{},\, \SI{0.4}{},\, \SI{1}{}$ and $\SI{2}{cm}$ in panels a) to d), respectively. The outline of the figures represents the arena boundary by which the ring's movement is confined. The opening at $(x, y) = (\SI{78}{cm}, \SI{0}{cm})$ represents the exit of the arena. For every size, a trajectory is partially highlighted, and a schematic ring illustrates the transported object. We have excluded the last $\SI{10}{cm}$ ($x > \SI{68}{cm}$) from our statistical analysis because strong deviations from the central line ($y \gg 0$) cause the preferred angle pointing towards the exit to strongly deviate from $\theta_0 = 0$ close to the exit.
			The two-dimensional velocity heat maps for the experimental trails are shown in panels e) to h). The velocity is biased in the target direction $\theta_0=0^\circ$ and centered around a preferred magnitude $v_c$ with width $\alpha_v v_c$. Most notably, the velocities $v_x$ and $v_y$ along the principal axes are not independent. With increasing load radius, the histograms become more concentrated around the mean magnitude and the target direction $\vec{e}_x$, which is also illustrated in Fig.~\refsub{fig:3}{a-b}. Sections of the trails in which the load did move for more than $\SI{10}{s}$ because of momentary dropping of the load (i.e., no ants were attached to the ring) where excluded when plotting the velocity histograms.
			Simulated trails and velocity heat maps are shown in panels i) to l) and m) to p), respectively. 
			The simulations are based on Eq.~\eqref{eq:langevin_r}-~\eqref{eq:langevin_t} with parameters taken from fitting the experimental velocity histograms to the theoretical distributions in Eqs.~\eqref{eq:angular_distribution}~-~\eqref{eq:speed_distribution} and shown in Fig.~\refsub{fig:3}{c,d}.}
		\label{fig:2}
	\end{figure*}
	
	\subsection{Experimental setup}
	In order to compare the trajectories of cooperative transport between two given points, an enclosed arena $(\SI{78}{cm} \times \SI{34}{cm})$ was placed $\sim \SI{1}{m}$ from a nest of \textit{Paratrechina longicornis} ants in the field as illustrated in Fig.~\refsub{fig:1}{a}. The arena had a $\sim \SI{1}{cm}$ wide single opening directed towards the nest. A load was repeatedly placed at a predetermined location on the far side of the arena. The initial load position, the arena entrance, and the nest entrance were all aligned (\textit{cf.}~supplementary material~\footnote{For further details on the experimental setup, the complete experimental data, further details of the analysis, numerical simulation results, and additional statistical tests, see Supplementary Material, which includes Refs.~\cite{McCreery2019, Martin2021, Amir2020} additional to those already mentioned here.}).
	
	Silicon rings of varying radii were incubated in cat food overnight in order to make them as attractive as food to the ants. A single ring of radius $r = \SI{0.15}{cm}, \SI{0.4}{cm}, \SI{1}{cm}$ or $\SI{2}{cm}$ was placed inside the arena, $\sim \SI{75}{cm}$ away from the arena entrance. After ants discovered the ring, they then cooperatively transported it towards the nest (Fig.~\refsub{fig:1}{b,c}). This constitutes a single trajectory of the load. Once the ants reached the arena opening, the same ring was returned to the initial location, and cooperative transport was immediately resumed. This was repeated for $N_{traj} = 43-55$ times for each of the four load radii.
	This process was recorded and the position of the center of the ring was extracted, as displayed in Fig.~\refsub{fig:2}{a-d}. 
	In the following, we will discuss the trajectory features of different ring radii. We emphasize here that radius could be used interchangeably with the number of ants participating in the cooperative transport $N$ because the latter increases linearly with $r$~\cite{Note1}. 
	Each set of trajectories of a given ring radius has a perpendicular spread to the line connecting the initial position of the load and the arena exit $\Delta y$ as previously shown in~\cite{Feinerman2018}. Some trajectories contain turnarounds (i.e. loops). The movement is reminiscent of a biased Brownian walk towards the nest. The source of the movement's bias is the ants' motivation to quickly carry the object to the arena exit and subsequently to their nest. The transporting group is directed by a pheromone trail laid by non-carrying ants~\cite{Fonio2016} and by newly arriving ants who influence the group by persistently tugging the object in the nest direction for~$\sim \SI{10}{s}$~\cite{Gelblum2015, Feinerman2017}.
	
	\subsection{Velocity distributions}
	Plotting the velocity distributions for sets of trajectories of varying radii has proven most indicative of the motion's nature (Fig.~\refsub{fig:2}{e-h}). The velocity magnitude $v$ adheres to a preferred value $v_c$ and exhibits concentric features. The angle of the velocity vector $\theta$ is strongly biased towards the nest direction ($\theta_0=0$).
	
	For smaller radii, $v$ is not as tightly bound to $v_c$ and \textcolor{black}{the angular spread} $\Delta\theta$ around $\theta_0$ is larger (Fig.~\refsub{fig:2}{e-f}). These trajectories are more erratic and display a wider perpendicular spread as shown in Fig.~\refsub{fig:2}{a, b}.
	For larger shapes, $v$ adheres more closely to $v_c$ and the uni-directionality towards the nest is more pronounced (Fig.~\refsub{fig:2}{g-h}). This corresponds to the trajectories' larger angular persistence and more narrow perpendicular spread as shown in Fig.~\refsub{fig:2}{c, d}. These observations are in agreement with previous empirical results~\cite{Gelblum2015}. In the following, we will present a self-propelled particle model that captures these features.
	
	\begin{figure*}[t]
		\centering
		\includegraphics[width=\linewidth]{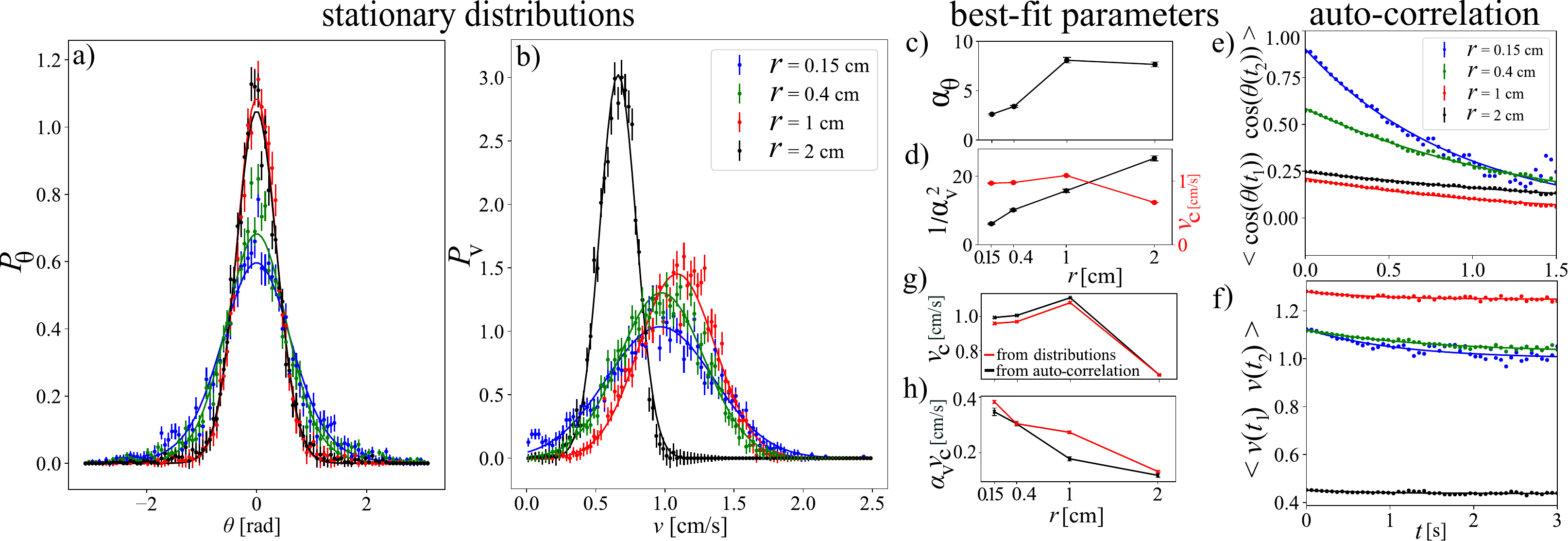}
		\caption{Stationary a) angular distributions and b) velocity magnitude distributions for different load radii. The dots correspond to experimental results. The solid lines represent the least square fit of the theoretical distributions, given in Eqs.~\ref{eq:angular_distribution} and \ref{eq:speed_distribution}.
			c) Best-fit parameters $\alpha_\theta$ for different radii resulting from the fitting of the angular distributions quantifying the strength of the returning force towards $\theta_0$.
			d) Best-fit parameters $1/\alpha_v^2$ and preferred velocity magnitude $v_c$ for different load radii, resulting from the fitting of the magnitude. $1/\alpha_v^2$ quantifies the strength of the returning force towards $v_c$ and increases with $r$. The preferred velocity magnitude has a maximum for the ring with $r = \SI{1}{cm}$ and drops significantly for the ring with radius $r = \SI{2}{cm}$.
			Auto-correlation of the time-series of e) $\cos(\theta(t))$ and f) velocity magnitude $v(t)$. The solid line represents the least square fit of the functions $g_{vv}(t)=\langle v(0) v(t)\rangle=v_c^2+\alpha_v v_c^2e^{-t/\tau_v}$ and $g_{\theta\theta}(t)\sim e^{-t/\tilde\tau_\theta}$. The auto-correlation of the velocity magnitude and angle decay over time with decay constants $\tau_v$ and $\tilde{\tau}_\theta$, respectively, which are shown in Fig.~\ref{fig:4}. Best-fit parameters g) $v_c$ and h) $\alpha_v v_c$ extracted by fitting velocity magnitude distribution $P_v$ (red) and the auto-correlation $g_{vv}(t)$ (black).}
		\label{fig:3}
	\end{figure*}
	
	
	\section{Model of ants as a single self-propelled particle}
	\subsection{Macroscopic trajectory features and model choice}
	While the stochastic motion of the rings crosses over into regular diffusion at late times, the velocity's angle $\theta$ and magnitude $v$ exhibit \textit{persistence} for short times below characteristic timescales $\tau_\theta$ and $\tau_v$, respectively. 
	Persistence implies that $v$ and $\theta$ are not completely randomized between time steps as would be the case in regular diffusion. Instead, the object partially preserves its previous velocity angle and magnitude. 
	Furthermore, the velocity components in $x$ and $y$ direction are not independently randomized. Instead, the velocity adheres to a preferred velocity magnitude $v_c$.
	One of the simplest implementations of this notion is to assume that $v$ itself performs a Brownian motion~\cite{Schienbein1993}. 
	In such a model, the self-propulsion is subject to friction, $v/v_c-1$, which is dimensionless and depends on the difference between the desired velocity magnitude $v_c$ and the instantaneous magnitude $v$. 
	For velocities $v<v_c$, this friction becomes negative, i.e., the self-propulsion works to increase $v$, and vice versa. Thus, the steady-state distribution is centered around $v_c>0$. The resulting self-propelled motion resembles an Ornstein-Uhlenbeck process~\cite{Elgeti2015,Bechinger2016} in velocity space. This has been discussed by Schienbein and Gruler~\cite{Schienbein1993} and others~\cite{Dunn1987, Alt1990}, and originally was used to describe the migration of single cells.
	One of the hallmarks of such a motion is that the independent degrees of freedom are the magnitude and the angle of the velocity vector.
	Based on our experimental findings, we thus describe the trajectories as a self-propelled motion with a velocity kernel.
	To this end, consider the Langevin equations for Brownian motion of the instantaneous velocity $\vec{v}$~\cite{Schienbein1993}:
	
	\begin{align}
		\label{eq:langevin_r}
		\dot{\vec{r}}&=\vec{v},\\
		{\dot v}
		&=-\tfrac{v_c}{\tau_v}(v/v_c-1) + \alpha_v v_c\sqrt{2/\tau_v}  \eta_v,
		\label{eq:langevin_v}\\
		\dot \theta&=-\tfrac{\alpha_{\theta}}{\tau_\theta}\sin(\theta-\theta_0)+
		\sqrt{2/\tau_\theta}\eta_\theta.
		\label{eq:langevin_t}
	\end{align}
	
	This set of equations describes the time evolution of the load's position $\vec{r}$ and its velocity $\vec{v}$ which is composed of its magnitude $v=\sqrt{v_x^2+v_y^2}$ and direction given by angle $\theta$, defined by $\tan\theta=v_y/v_x$. The normalized noise terms are delta-correlated, $\langle \eta_v(t)\eta_v(t')\rangle=
	\langle \eta_\theta(t)\eta_\theta(t')\rangle=\delta(t-t')$.
	The motion is characterized by $v$ and $\theta$ that relax towards their preferred values, $v_c$ and $\theta_0$. The relaxation of $v$ towards $v_c$ corresponds to the aforementioned dimensionless friction resulting in self-propulsion. Simultaneously, the motion is slowly randomized by the Gaussian stochastic variables $\eta_v$ and $\eta_\theta$
	~\footnote{We note that $\theta$ is periodic with $2\pi$, therefore it is an obvious choice in the Langevin equation for $\theta$ to employ a periodic rectification force using a sine function~\cite{Schienbein1993}.}. The spread of $v$ around $v_c$ due to randomization is quantified by $\alpha_v v_c$. 
	The timescale $\tau_v$ is the time it takes for velocities $v\neq v_c$ to decay towards $v_c$ while $\tau_\theta$ is the time it takes for angles $\theta\neq \theta_0$ to decay towards $\theta_0$.
	The long-range attraction to the nest is encoded by a preferred velocity direction $\theta_0$, with strength $\alpha_{\theta}$.
	A subtlety of the model presented in Eqs.~(\ref{eq:langevin_r}-\ref{eq:langevin_t}) is that $v$ may formally acquire negative values.
	However, in the experimental data, only very few points are close to $v=0$. We can, therefore, safely disregard such events in the following.
	It is also possible - if somewhat cumbersome - to include a local potential per mass unit $V(\vec{r})$, which creates an additional, spatially dependent acceleration $-\vec{\nabla} V(\vec{r})$ acting on $\dot{\vec{v}}$. This could be used to capture the effect of potential pheromone trails guiding the load away from the preferred direction of motion $\theta_0$. Such are not present in our case and can safely therefore be disregarded for the following analysis~\cite{Note1}. 
	
	\subsection{Possible turning mechanisms}
	
	The model contains two velocities $v_c$, $\alpha_v v_c$, and two timescales $\tau_v$, $\tau_\theta$. 
	The dimensionless parameter $\alpha_{\theta}$ measures the strength of the global bias, i.e. how closely the target direction $\theta_0$ is adhered to. The motion becomes increasingly unidirectional the larger $\alpha_{\theta}$ is, while for $\alpha_{\theta}=0$, the motion is isotropic.
	Similarly, the dimensionless ratio $\alpha_v$ encodes 
	how strongly the velocity magnitude is randomized, causing $v$ to deviate from $v_c$.
	If $\alpha_v > 1$, the velocity is randomized so quickly that $\theta$ can flip in a short time merely due to the large magnitude fluctuations.
	In contrast, if $\alpha_v< 1$, $v$ adheres close enough to $v_c$, such that the velocity does not flip its angle. Then turnaround (i.e., a reversal of the velocity vector) only happens by slowly rotating $\theta$ while moving at nonzero velocity. 
	For this latter case, one can distinguish whether the trajectory turns around more quickly due to the randomization of the angle or due to the influence of the bias: Consider a point in the trajectory where the object moves opposite to the global bias. Following the factors preceding the restoring and randomizing components in Eq. \eqref{eq:langevin_t}, turning around due to angular fluctuations happens on a time scale of $T_{r}=\tau_\theta$, while the turnaround time towards the nest due to the bias is $T_{b}=\tau_\theta/\alpha_\theta$. By observing a turnaround, we cannot distinguish whether it occurred due to randomization of the angle or the global bias. However, from experiments, we can extract an effective angular correlation time $\tilde\tau_\theta=\tilde\tau_\theta(\tau_\theta,\alpha_\theta)$, which takes into account both turning mechanisms and can be analytically approximated~\cite{Note1}. Therefore, given that we determined $\alpha_\theta$ and $\tilde\tau_\theta$ from a set of experimental trajectories, we are able to approximate $\tau_\theta$. 
	
	\section{Estimation of model parameters}
	All model parameters are accessible by fitting the experimental data: $\alpha_v$ and $\alpha_{\theta}$ can be extracted from the steady state probability distribution in velocity magnitude and angle. The timescales $\tau_v$ and $\tau_\theta$ follow from their respective autocorrelation functions.
	
	\subsection{Parameters from velocity distributions}
	The distributions of the measured velocity components in angle and magnitude are shown in Fig.~\refsub{fig:3}{a} and~\refsub{fig:3}{b}. 
	They are well described by the steady-state solutions of Eqs.~(\ref{eq:langevin_v}-\ref{eq:langevin_t}) for $\theta$ and $v$,
	\begin{align}
		P_\theta&=C_\theta^{-1} \exp{\bigl(\alpha_{\theta}\cos(\theta-\theta_0) \bigr)},
		\label{eq:angular_distribution}\\
		P_v&=C_v^{-1}
		\exp{\Bigl(-\frac{(1-v/v_c)^2}{2\alpha_v^2}\Bigr)}.
		\label{eq:speed_distribution}
	\end{align}
	where $C_\theta=2\pi I_0(\alpha_{\theta})$, 
	$C_v=\sqrt{2\pi}\alpha_vv_c$
	are normalization constants, and $I_0$ is the modified Bessel function of the first kind. 
	The bias strength $\alpha_{\theta}$ was determined by fitting $P_\theta$ to the angular distributions (Fig.~\refsub{fig:3}{a}), $\alpha_v$ and $v_c$ were determined by fitting $P_v$ to the velocity magnitude distributions (Fig.~\refsub{fig:3}{b}).
	After normalizing velocity $\vec{v}$ by $v_c$, its steady state properties only depend on $\alpha_v$ and $\alpha_{\theta}$.
	The best-fit values for these latter two parameters are shown in Fig.~\refsub{fig:3}{c, d} and are a measure for the strength of the returning force towards $v_c$ and $\theta_0$.
	We find that both $1/\alpha_v^2$ and $\alpha_{\theta}$ increase with increasing ring radius, which is consistent with the trajectories' stronger adherence to the target values $(v_c,\theta_0)$ with increasing load radius.
	The slight increase in $P_\theta$ at $v\approx 0$ for $r=\SI{0.15}{cm}$ in our experimental data presented in Fig.~\refsub{fig:3}{b} arises from temporary dropping of the load, which the fitted distribution does not capture.

	\begin{figure*}
		\centering
		\includegraphics[width=0.85\linewidth]{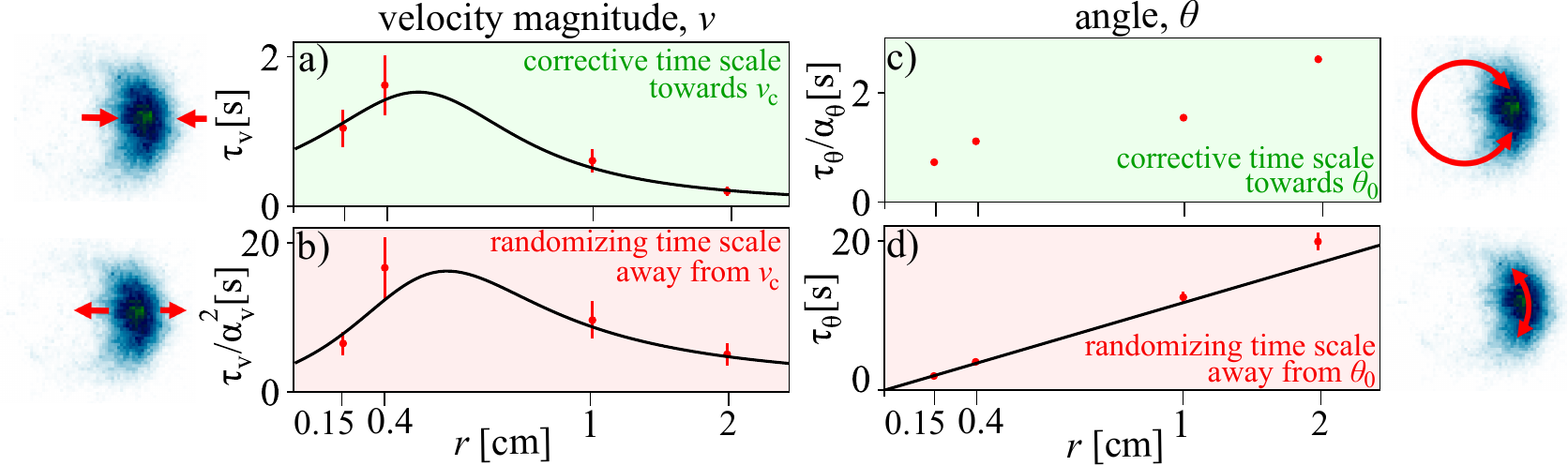}
		\caption{
			Optimal fit parameters taken from angular and velocity magnitude distributions and from the autocorrelation as a function of $r$. These give time scales of the corrective and randomizing force on the (a-b) velocity magnitude and (c-d) angle. The black curves represent the best fit of the phenomenological scaling ansatz with parameters $\Delta$, $N_0$, and $\tau_0$. No scaling relation was suggested for $\alpha_{\theta}$; therefore, c) contains no curve. On the left and right are depictions of the respective effect of each force term on the respective distributions.
		}
		\label{fig:4}
	\end{figure*}
	
	\subsection{Parameters from dynamical trajectory features}
	To experimentally determine the dynamical parameters $\tau_v$ and $\tau_\theta$, we turn towards the autocorrelation functions. The correlation function for the velocity magnitude is known exactly~\cite{Schienbein1993}, and reads $g_{vv}(t)=\langle v(0) v(t)\rangle=v_c^2+(\alpha_v v_c)^2e^{-t/\tau_v}$. Our experimental results and the least square fit of $g_{vv}$ are shown in Fig.~\refsub{fig:3}{f}.
	The optimal fit parameters $v_c$ and $\alpha_v v_c$ take on similar values to the ones found from the stationary velocity distributions as shown in Fig.~\refsub{fig:3}{g, h}.
	In both cases, $v_c$ takes on a maximal value at ring radius $r=\SI{1}{cm}$, and $\alpha_v v_c$ decreases monotonically with ring radius. The steady-state and the dynamical properties are the result of the same dynamical relations.
	Since the Langevin Eq.~\eqref{eq:langevin_t} for the angle is nonlinear, the time-dependent cosine-correlation function $g_{\theta\theta}(t)=\langle \cos \theta(0) \cos\theta(t)\rangle$ has to be calculated perturbatively in $\alpha_{\theta}$~\cite{Note1}. We find that the exponential decay of the correlations, $g_{\theta\theta}(t)\sim e^{-t/\tilde\tau_\theta}$,
	contains the effective decay rate 
	$\tilde\tau_\theta^{-1}\approx\tau_\theta^{-1}(1+\alpha_{\theta}^2/6+\dots)$.
	Performing a fit for the autocorrelation functions at short times~\cite{Note1} yields $\tau_v$ and $\tau_\theta$, which are shown in Fig.~\refsub{fig:4}{a, d}, respectively. For all radii, $T_r > T_b$, from which we conclude that an object moving away from the nest turns around more quickly due to the global bias, not the randomization of the angle. From the experimentally determined turning time $\tilde\tau_\theta$ and preferred velocity magnitude $v_c$, one can calculate the turning radius $R=v_c\tilde\tau_\theta$. We find that $R=\SI{0.7}{}-\SI{1.7}{cm}$, which is in line with previously reported values for the turning radius~\cite{Feinerman2018}.
	
	\subsection{Simulations confirm proper model choice}
	Based on the Langevin equations~(\ref{eq:langevin_r}-\ref{eq:langevin_t}) and best-fit parameters (Fig.~\refsub{fig:3}{c, d}), we numerically simulated 10 trajectories per ring radius. We use $\Delta t = \SI{0.1}{s}$ and simulated the trajectories until they reached $x > \SI{78}{cm}$, which took approximately $10.000$ time steps per trajectory. These trajectories and their respective velocity histograms are shown in Fig.~\refsub{fig:2}{i-p}. 
	The simulated and experimental trajectories show similar properties (Fig.~\ref{fig:2}): The trajectories of smaller ring radii are more erratic, while those of larger ring radii are more smooth. Also, we find strong agreement between simulated and experimental velocity heat maps (Fig.~\refsub{fig:2}{m-p}). This further validates our self-propelled particle model's accuracy in capturing the load behavior and confirms the correctness of our chosen model parameters.
	
	We find that the spread of the experimental trails $\Delta y$ remains uniform with growing $x$, while for simulated trajectories, the $\Delta y$ grows. This is due to the fact that in experiments, the load is biased towards the exit door of the arena, not merely $\theta_0$. Incorporating this spatial dependency into Eq.~\eqref{eq:langevin_t} leaves the velocity histograms mainly unaffected because during most of the trail the arena exit lies at an angle $\approx \theta_0$. Consequently, we choose not to include this spacial dependency in our model.
	
	\section{Scaling analysis}
	\subsection{Comparing dynamics of varied load sizes}
	We now investigate the dependence of the model parameters on load radius, with the goal of finding a meaningful connection between the dynamics of the effective self-propelled particle and individual ant behavior.
	To this end, given both steady state parameters $(\alpha_v^{-2},\alpha_{\theta})$ and the dynamical parameters ($\tau_v$, $\tau_\theta$) for all four load radii, we analyze each of the four force terms appearing in Eqs.~(\ref{eq:langevin_v}-\ref{eq:langevin_t}) separately: The deterministic relaxation towards $(v_c,\theta_0)$, encoded by the friction rates $(1/\tau_v,\alpha_{\theta}/\tau_\theta)$, as well as the rates of randomization $(\alpha_v^2/\tau_v,1/\tau_\theta)$.
	The resulting characteristic time scales as a function of $r$ are shown in Fig.~\ref{fig:4}.
	The angular rates ($\tau_\theta$, $\alpha_{\theta}/\tau_\theta$) increase monotonically with increasing ring radius (Fig.~\refsub{fig:4}{b, d}), which is consistent with previous research showing a decrease in trajectory curvature with increasing radius~\cite{Gelblum2015}.
	On the other hand, $\tau_v$ and $\tau_v/\alpha_v^2$ has a maximum at an intermediate ring radius of $r=\SI{0.4}{cm}$ (Fig.~\refsub{fig:4}{a, c}). Previously, using the coupled-carriers model a phase transition at this load radius was identified~\cite{Gelblum2015}. It is known that close to a phase transition, the recovery rate after small perturbations slows down, leading to an increase in the temporal autocorrelation of the magnetization, a phenomenon known as 'critical slowing down'~\cite{Scheffer2009}. In our experimental system, the load's velocity magnitude is determined by the internal alignment of the ants, which can be thought of as the system's magnetization. Therefore, the increase in $\tau_\theta$ aligns with the previously identified critical point. 
	
	\subsection{Relating model parameters to our system of carrying ants}
	
	Next, we rationalize a simple scaling ansatz that leads to a maximal persistence time $\tau_v$ for intermediate group sizes.
	
	We assume that a small number of attached ants struggle to maintain the average velocity magnitude when repositioning or changing their state from lifting to pulling the load~\cite{Gelblum2015}. Therefore, the auto-correlation time $\tau_v$ is diminished. On the other hand, for a large number of ants, the target velocity $v_c$ is quickly reattained by averaging if a disturbance made it deviate. 
	In between these two limits, for carrier numbers in the range of 10-15, the ants keep a once acquired velocity for a maximal time $\tau_0$. Therefore, we propose the following scaling relation:
	
	\begin{equation}\label{eq:model1}
		\frac{1}{\tau_v}=\frac{(N-N_0)^2+\Delta^2}{\Delta^2\tau_0}.
	\end{equation}
	
	Here, $N_0$ is the number of ants at which the effective friction coefficient is minimal, while $\Delta$ is an estimate of the fluctuation in the number of attached ants. While the ansatz chosen here is probably not unique, the location and width of the maximum in $\tau_v$ and the typical timescales in the experiment strongly insinuate that comparable values would be recovered in more elaborate scenarios.
	
	In contrast to the scaling of $\tau_v$, the angular relaxation rate $\tau_\theta$ increases essentially linearly with $r$. 
	This makes sense if the travel direction $\theta$ is collectively negotiated~\cite{Feinerman2018}: 
	Assuming that the retention time for an individual ant is $\tau_0$, the collective memory about the target direction is averaged out (i.e., deleted) only after a much longer time $\tau_\theta\sim N\tau_0$~\cite{Boczkowski2018}:
	Therefore, in contrast to mechanically enforced compromises (tug-of-war), the travel direction - as a decision-based group effort -
	is not subject to the diminishing relative contribution of the individual.
	
	\begin{equation}\label{eq:model2}
		\frac{1}{\tau_\theta}=\frac{\pi}{ N\tau_0}
	\end{equation}
	Note that $\tau_\theta$, compared to $\tau_v$ is larger by a factor of $\pi$ due to the different normalization between $v$ and $\theta$.
	
	Regarding the normalized standard deviation in the velocity $\alpha_v$, we reiterate that the average velocity magnitude is not negotiated but emerges by averaging. Thus the ensemble average is expected to scale with a standard deviation $\alpha_v\sim N^{-1/2}$ according to the law of large numbers. 
	It is known experimentally that the target velocity is enforced by the pulling ants, which constitute around half of the attached ants~\cite{Gelblum2015}. Using this information, the scaling form Eq.~\eqref{eq:model1} is constructed such as to capture both the small-$N$ limit with $\alpha_v(N\to 0)=\Delta/2$ as well as the large-$N$ limit where $\alpha_v(N\to\infty)= N/2$.
	
	\begin{equation}\label{eq:model4}
		\alpha_v^{-2}=\tfrac{1}{2}(N+\Delta)
	\end{equation}
	Finally, we point out that the acceleration bias $\tau_\theta/\alpha_\theta$ (Fig.~\refsub{fig:4}{b}), does not lend itself to such a simple analysis. This is not unexpected because the strength of the bias $\alpha_{\theta}$ depends on the arrival rate of new, informed ants. Therefore, the timescale $\tau_\theta/\alpha_{\theta}$ is subject to environmental conditions which are not experimentally controllable in the present experiment which prevents us from to drawing firm conclusions. 
	The simultaneous fit yields the following parameter values, where error propagation was taken into account:
	\begin{align}
		N_0=10.06 \pm 1.67\\
		\Delta=9.95 \pm 0.32\\
		\tau_0=\SI{1.53}{s} \pm \SI{0.04}{s}
	\end{align}
	The goodness of fit qualifiers are $\chi^2=38.81$, $\text{objective value}=19.40$ and $r^2=0.980$.
	The fits are shown in Fig.~\ref{fig:4}a, c and d.
	
	Similar properties have been reported in Ref.~\cite{Gelblum2015} within numerical calculations within the coupled-carriers model. Namely, the timescale for reorientation of an ant was found to be $\tau_0^{\mathrm{lit}}\sim \SI{1.4}{s}$, and the most cooperative and responsive group sizes were in the range of $N_0^{\mathrm{lit}}\sim 10$ ants~\cite{Feinerman2018} corresponding to the aforementioned phase transition. 
	
	\section{Conclusions and outlook}
	We have modeled the movement of a load being cooperatively transported by ants as a biased self-propelled particle subject to velocity diffusion. We also related the presented statistical model to other microscopic features of the previously established coupled-carriers model using a simple scaling ansatz of the model parameters.
	We analyzed the key dynamical and steady-state properties of ant cooperative transport and found that the autocorrelation time of the velocity magnitude reaches a maximum for intermediate group sizes. We have found that this model can successfully reveal the same phase transition previously identified using the agent-based coupled-carriers model.
	However, the statistical approach presented in this study does not rely on specific knowledge about the behavior of individual ants. Instead, it uses a model describing a single self-propelled particle to reconstruct group behavior while making no assumptions about the individual agents. Therefore, we have shown that these types of models can provide a valuable tool for describing and understanding collective behavior in which microscopic details concerning the agents' individual and cooperative behavior are unknown. Furthermore, our findings open up the exciting possibility of using Ising-type models similar to the coupled-carriers model to analyze the behavior of individual, self-propelled agents.
	
	\begin{acknowledgments}
		We thank 
		Uzy Smilansky, Nir Gov, Harish Charan and David Mukamel
		for useful discussions. We thank Guy Han for the technical support. 
		O.~F. is supported by the Israel Science Foundation, Grant No. 1727/20, the Minerva Foundation, and the European Research Council (ERC) under the European Unions Horizon 2020 research and innovation program (Grant Agreement No. 770964).
	\end{acknowledgments}
	\bibliography{literature}
\end{document}